\documentclass[amssymb,prl,twocolumn,showpacs]{revtex4}
\usepackage{epsfig}
\usepackage{dcolumn}
\usepackage{amsmath}
\begin{document}
%\documentclass[12 pt,a4paper]{article} %selecciona el tipo de
%documento
%\usepackage[english]{babel} %selecciona el idioma
%\frenchspacing %trata los espacios despues de los puntos igual
%que los kotos
%\usepackage{epsfig}
%\usepackage{amsmath}
%\usepackage[a4paper,dvips]{geometry}
%\geometry{textwidth=16 cm, textheight=22 cm}
%\begin{document}
\title{\bf Electronic transport through a double quantum dot in the spin blockade regime:
Theoretical models}
\author{Jes\'us I\~narrea$^{1,2}$, Gloria Platero$^2$ and Allan H.
MacDonald$^3$} \affiliation {$^1$Escuela Polit\'ecnica
Superior,Universidad Carlos III,Leganes,Madrid,Spain\\ $^2$Instituto
de Ciencia de Materiales, CSIC,
Cantoblanco,Madrid,28049,Spain.\\
 $^3$Department of Physics,
University of Texas at Austin.Austin, Texas 78712}
\date{\today}
%%%%%%%%%%%%%%%%%%%%%%%%%%%%%%%%%%%%%%%%%%%%%%%%%%%%%%%%%%%%%%%%%%
%%%%%%%%%%%%
\begin{abstract}
We analyzed the electronic transport through a double quantum dot in
the spin blockade regime. Experiments of current rectification by
Pauli exclusion principle in double quantum dots were discussed. The
electron and nuclei spin dynamics and their interplay due to the
Hyperfine interaction were self-consistently analyzed within the
framework of rate equations. Our results show that the current
leakage experimentally observed in the spin-blockade region, is due
to spin-flip processes induced by Hyperfine interaction through
Overhauser effect. We show as well how a magnetic field applied
parallel to the current allows excited states to participate in the
electronic current and removes spin blockade. Our model includes
also a self-consistent description of inelastic transitions where
the energy is exchanged through interactions with acoustic phonons
in the environment. It accounts for spontaneous emission of phonons
which results in additional features in the current characteristics.
We develop a microscopical model to treat the Hyperfine interaction
in each dot. Using this model we study the dynamical nuclear
polarization as a function of the applied voltage.
\end{abstract}
%%%%%%%%%%%%%%%%%%%%%%%%%%%%%%%%%%%%%%%%%%%%%%%%%%%%%%%%%%%%%%%%%%
%%%%%%%%%%%%
\maketitle
\section{I. Introduction}
Recent transport experiments in vertical double quantum dots (DQD's)
show that Pauli exclusion principle is important\cite{ono} in
current rectification. In particular, spin blockade (SB) is observed
at certain regions of dc voltages. The interplay between Coulomb and
SB can be used to block the current in one direction of bias while
allowing it to flow in the opposite one. Johnson et
al.\cite{johnson} observed as well SB and spin rectification in a
lateral DQD. Then DQD's could behave as externally controllable
spin-Coulomb rectifiers with potential application in spintronics as
spin memories and transistors.

Spin de-coherence\cite{fujisawa,gymat} and relaxation processes
induced, for instance, by spin-orbit (SO) scattering \cite{golovach}
or Hyperfine (HF) interaction \cite{erlingsson}, have shown to
reduce SB producing a leakage current. In this paper we
theoretically analyze recent experiments of transport through two
weakly coupled QD's \cite{ono}. We consider simultaneously Hyperfine
(HF) interaction and emission of phonons to be responsible of the SB
lifting and the main features in the experimental current/voltage
($I/V_{DC}$) curve. According to our calculations, inter-dot
phonon-assisted tunneling has to be included to have a full
understanding of the physics behind the experimental results. In the
corresponding experiment\cite{ono}, the total electron number of the
system is fluctuating between one and two. Due to the different
gates voltages applied between the two dots the left dot $(n_1)$ can
have up to one electron and the right one $(n_2)$ can fluctuate
between one and two keeping the sum $(n_1+n_2)$ between one and two.
Current flow is allowed when the electrons in each QD have
antiparallel spins and a finite gate voltage allows one electron in
the left dot to tunnel sequentially to the right one and further to
the collector. However, for weakly coupled QD's there is a similar
probability for the electron coming from the left lead to be
parallel or antiparallel to the electron spin occupying the right
dot. In the first case, the electron cannot tunnel to the right dot
due to Pauli exclusion principle and SB takes place, presenting a
plateau in the $I/V_{DC}$ curve.

The theoretical model presented in this paper has been developed in
the frame of rate equations. We solve self-consistently a system of
coupled time-evolution equations for electronic charge occupations
and nuclei polarizations. Our theoretical results reproduce the
$I/V_{DC}$ observed plateau due to SB and also the main current
peaks. HF interaction is proposed as the candidate to lift SB,
producing spin-flip (sf) of electrons and nuclei. On the other hand
phonon-assisted tunneling is proposed, in parallel with the direct
tunneling, to sustain the total current through the device. The
electrons and nuclei spin interactions brings to the Overhauser
effect, which is also called flip-flop interaction because each time
the electron flip the spin up to down (down to up) the nuclear spin
does the opposite. According to measurements on QD's by Fujisawa et
al.\cite{fujisawa} the spin-flip time, $\tau_{sf}>10^{-6}$s, is much
longer than the typical tunneling time, $\tau_{tun}=1-100$ ns, or
the momentum relaxation time, $\tau_{mo}=1-10$ ns, meaning that
spin-flip processes due to HF interaction are important mostly in
the SB region. Our system consists of a vertical DQD under an
external DC voltage in the presence of a magnetic field parallel to
the current.

\section{II. Theoretical model}
\subsection{Elastic and inelastic tunneling}
We consider a hamiltonian: $H=
H_{L}+H_{R}+H_{T}^{LR}+H_{leads}+H_{T}^{l,D}$ where $ H_{L} (H_{R})$
is the hamiltonian for the isolated left (right) QD and is modelled
as one-level (two-level) Anderson impurity.
$H_{T}^{LR}$($H_{T}^{l,D}$) describes tunneling between QD's (leads
and QD's)\cite{cota}) and $H_{leads}$ is the leads hamiltonian. We
use a basis that contains 20 states given by:\\
\begin{eqnarray}
&&|1\rangle=|0,\uparrow\rangle;|2\rangle=|0,\downarrow\rangle;
|3\rangle=|\uparrow,\uparrow\rangle;
|4\rangle=|\downarrow,\downarrow\rangle;\nonumber\\
&&|5\rangle=|\uparrow,\downarrow\rangle;
|6\rangle=|\downarrow,\uparrow\rangle;|7\rangle =
|0,\uparrow{\uparrow}^*\rangle;
|8\rangle=|0,\downarrow{\downarrow}^*\rangle;\nonumber\\
&&|9\rangle=|0,\uparrow\downarrow\rangle; |10\rangle=
|\uparrow,0\rangle; |11\rangle=|\downarrow,0\rangle; |12\rangle
=|0,\uparrow^*\rangle; \nonumber\\
&&|13\rangle=|0,{\downarrow}^*\rangle;
|14\rangle=|\uparrow,{\uparrow}^*\rangle; |15\rangle=
|\downarrow,{\downarrow}^*\rangle;\nonumber\\
&&|16\rangle=|\uparrow,{\downarrow}^*\rangle;
|17\rangle=|\downarrow,{\uparrow}^*\rangle;
|18\rangle=|0,0\rangle;\nonumber\\
&&|19\rangle=|0,\downarrow,{\uparrow}^*\rangle;
|20\rangle=|0,\uparrow{\downarrow}^*\rangle\nonumber
\end{eqnarray}

 We have considered two
levels in the right QD. Those states marked with (*) correspond to
the excited state in the right QD. Double occupied states in the
left QD do not participate in the electron transport except for high
reverse bias. For simplicity we do not consider here this regime.
The system of equations of motion for state occupation probabilities
$\rho_{s}$ includes scattering in and scattering out
contributions\cite{blum,mahler,loss,cota}:
\begin{equation}
\dot{\rho}(t)_{s} = \sum_{m\neq s} W_{sm}\rho_{m} - \sum_{k\neq s}
W_{ks}\rho_{s}
\end{equation}
Our neglect of coherence\cite{cota} effects is appropriate for
weakly coupled quantum dots. $W_{i,j}$ is the transition rate state
j to state i.

For the contact-QD tunneling rate $W_{i,j}$ we use a Fermi Golden
Rule (FGR) expression, i.e., first order time-dependent perturbation
theory \cite{dong,wiel,gurvitz}. The expression for the left
contact-QD tunneling reads:
\begin{equation}
W_{L,1(1,L)} =\frac{\pi}{\hbar} T^2\rho \left[ \frac{1}{2} +(-)
\frac {1}{\pi}arctg\left(
\frac{\mu_L-\mu_1+eV_{L1}}{\gamma}\right)\right]
\end{equation}
where $W_{L,1(1,L)}$ is the tunneling probability for the left
contact (QD) to the left QD (contact) . A similar expression can be
obtained for the the right QD to the right contact:
\begin{equation}
W_{R,2(2,R)} =\frac{\pi}{\hbar} T^2\rho \left[ \frac{1}{2} +(-)
\frac {1}{\pi}arctg\left(
\frac{-\mu_R+\mu_2-eV_{2R}}{\gamma}\right)\right]
\end{equation}
$T$ is the transmission amplitude of the outer barriers,
($T\simeq1.3\times10^{-3}meV$). $\gamma$ is the width of the QD
state.
%Lorentzian-shape therefore corresponding
%to the situation where elastic scattering is important only. Thus
%$\gamma$ is of the order of the transmission through the device
%barriers, i.e., $\gamma \sim \mu eV$ .
$\mu_{L,(R)}$ is the chemical potential in the left contact (right)
($\mu_{L}=4 meV$, $\mu_{R}=4 meV$ for zero bias\cite{ono}),
$\mu_{1,(2)}$ is the chemical potential in the left dot (right dot).
To calculate $\mu_{1,(2)}$ we have used, apart from the
corresponding voltage drops through the device, the intradot and
interdot charging energies: $4$ and $2$ meV respectively\cite{ono}.
$\rho$ the two-dimensional density of states, $V_{L1}$ is the
potential drop between the emitter (left contact) and the left QD
and $V_{2R}$ is the potential drop between the right dot and the
collector (right contact).

Inter-dot transition rates account for both elastic tunneling and
inelastic phonon assisted tunneling. The corresponding expression
for the elastic inter-dot tunneling is given by:
\begin{equation}
W_{1,2(2,1)} = \frac{T_{1,2}^2}{\hbar} \left[
\frac{\gamma}{(\mu_1-\mu_2+eV_{12})^{2}+\gamma^{2}}\right]
\end{equation}
where  $T_{1,2}$ is the transmission through the inner barrier
($T_{1,2}\simeq5\times10^{-2}meV$) and $V_{12}$ is the voltage drop
between the QD's.
\begin{figure}
%\centering\epsfxsize=3.6in \epsfysize=4.2in \epsffile{figure1.eps}
\centering\epsfxsize=3.6in \epsfysize=2.0in \epsffile{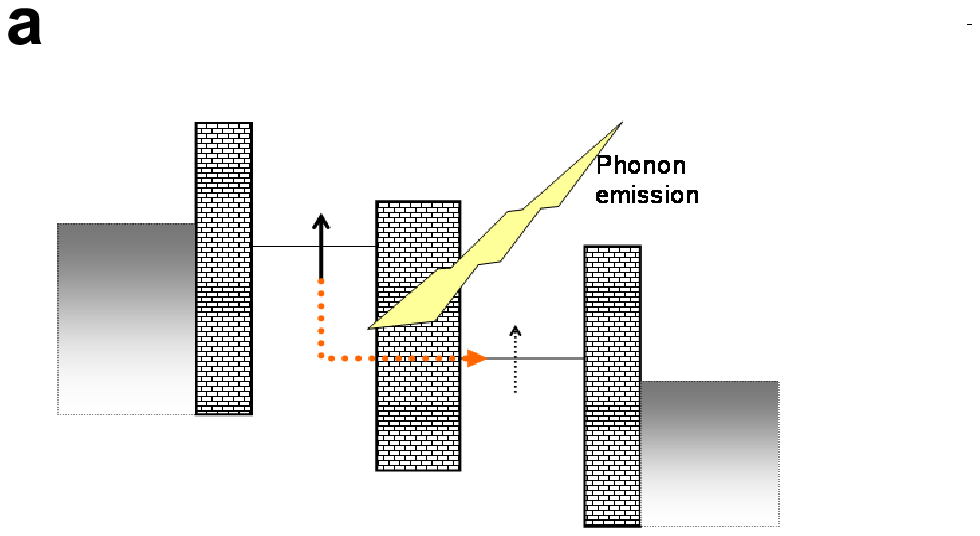}
\centering\epsfxsize=3.6in \epsfysize=2.0in \epsffile{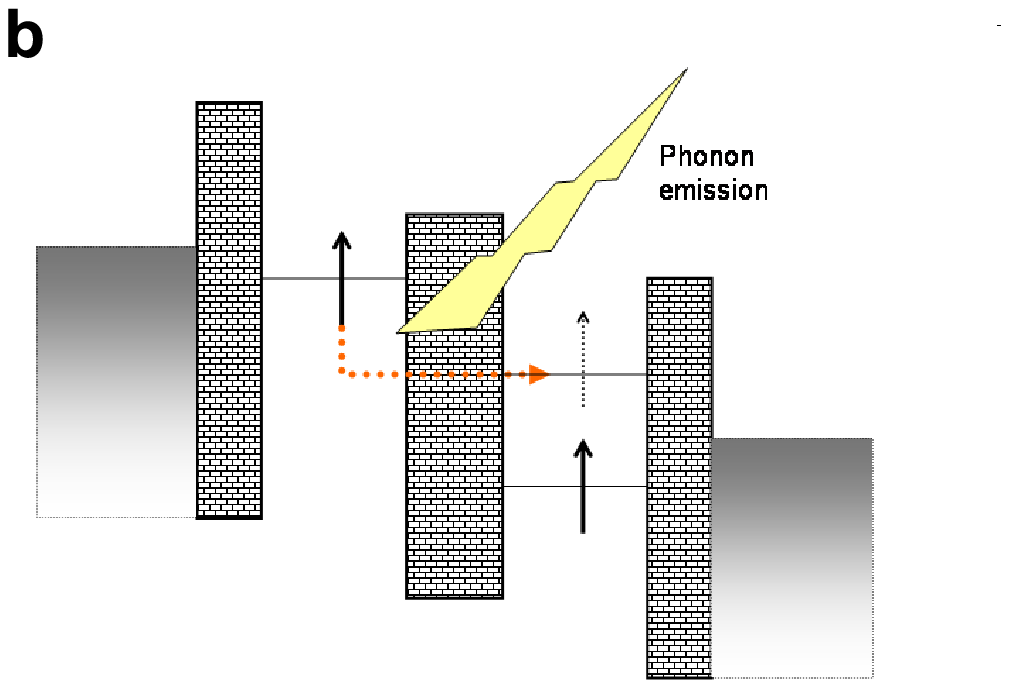}
\caption{(Color on line). Schematic diagrams for inelastic
phonon-assisted tunneling between weakly coupled dots. a) Between
ground levels in each QD. b) Between the ground level of the left QD
and the first excited level of the right QD.}
\end{figure}

For inelastic transitions, energy is exchanged with phonons in the
environment. In other words, at $T\approx 0$ (we have considered
zero temperature in our calculations) the inelastic tunneling
between the two dots is assisted by the emission of acoustic
phonons, yielding a significant contribution to the current. This
contribution has been experimentally measured by Fujisawa et
al.,\cite{fujisawa2} and theoretically analyzed by Brandes et
al.,\cite{brandes}. In order to calculate the inelastic transition
rate $W^{ph}_{1,2}$ due to the emission of phonons, we have
considered the theory developed by Brandes et al.,\cite{brandes}.
Including piezoelectric and deformation potential acoustic phonons
the transition rate reads:

\begin{equation}
W^{ph}_{1,2} = \frac{\pi T_{12}^2}{\hbar} \left[
\frac{\alpha_{pie}}{\varepsilon}+
\frac{\varepsilon}{\hbar^{2}w^{2}_{\xi}}\right]\left[1-\frac{w_{d}}{w}\sin\frac{w}{w_{d}}\right]
\end{equation}

where $\alpha_{pie}$ is a piezoelectric coupling parameter
($\alpha_{pie}=0.02$), $\varepsilon=\hbar
w=\mu_{1}-\mu_{2}+eV_{12}$, $w_{d}=c/d$ being $c$ the sound velocity
and $d$ the distance between the dots. Finally,
\begin{equation}
\frac{1}{w^{2}_{\xi}}=\frac{1}{\pi^{2}c^{3}}\frac{\Xi^{2}}{2\rho_{M}c^{2}\hbar}
\end{equation}
where $\rho_{M}$ is the mass density and $\Xi$ is the deformation
potential ($\Xi\simeq 7 eV$ for GaAs). In Fig. 1, we represent
schematically the inelastic contribution to  $I$ through the
emission of phonons, between the corresponding levels of each QD.

\subsection{Microscopical model for Hyperfine interaction}

We calculate the electronic spin-flip scattering rate $W_{i,j}^{sf}$
using a microscopic model that accounts for HF interactions and
external magnetic fields:
\begin{equation}
 \hat{H}=g_{e}\mu_{B}\vec{S\cdot}\vec{B}+
\frac{A}{N_{L(R)}}\sum_{i=1}^{N_{L(R)}}\left[S_{z}I_{z}^{i}+
\frac{1}{2}(S_{+}I_{-}^{i}+S_{-}I_{+}^{i})\right]
\end{equation}
where $A$ is the average HF coupling constant, ($A=90\mu eV$ for
GaAs \cite{paget}) and $I$ the nuclear spin . $N_{L(R)}$ is the
number of nuclei in the left (right) dot, ($N_{L}=10^{6}$ and
$N_{R}=1.1\times10^{6}$). For simplicity we assume that $I=1/2$. We
take $B$ to be oriented along the $\hat{z}$ direction (current
direction). The HF interaction can then be separated into mean-field
and flip-flop contributions:
\begin{equation}
 \hat{H}= \hat{H}_{z}+\hat{H}_{sf}
\end{equation}
 where
\begin{equation}
 \hat{H}_{z}= [ g_{e}\mu_{B} B  + A \langle I_{z} \rangle_{L(R)} ] S_z
\end{equation}
being,
\begin{eqnarray}
\langle I_{z} \rangle_{L(R)}&=&
\frac{1}{N_{L(R)}}\sum_{i=1}^{N_{L(R)}}(I_{z}^{i})_{L(R)}\nonumber\\
&=&\left[\frac{N^\uparrow-N^\downarrow}{N^\uparrow+N^\downarrow}\right]_{L(R)}|I_{z}|\nonumber\\
&=&P_{L(R)}|I_{z}|
\end{eqnarray}
$P_{L(R)}=\left[\frac{N^\uparrow-N^\downarrow}{N^\uparrow+N^\downarrow}\right]_{L(R)}$
is the nuclear spin polarization where $N^{\uparrow(\downarrow)}$ is
the number of nuclei with spin up(down), in a QD. We have chosen
that initially, the nuclei polarization of the left and right dots
are equal to zero.

$\hat{H}_{z}$ has external and effective nuclear field
contributions. The latter
 given by:
\begin{equation}
B_{nuc} =\frac{A \langle I_{z} \rangle_{L(R)}}{g_{e} \mu_{B}}
\end{equation}

On the other hand:
\begin{equation}
\hat{H}_{sf}= \frac{A}{2N_{L(R)}}\sum_{i} \left[
S_{+}I_{-}^{i}+S_{-}I_{+}^{i}\right]
\end{equation}
is the flip-flop interaction responsible for mutual electronic and
nuclear spin flips. Nuclei in similar quantum dots can give rise to
different effective nuclear fields. This can be related with
different effective Hyperfine interactions in each dot. An slightly
distinct number of nuclei can explain the different Hyperfine
interactions and eventually the independent behavior in terms of the
effective nuclear fields. Our model accounts for this situation with
$N_{L(R)}$ and $\langle I_{z} \rangle_{L(R)}$.
\begin{figure}
\centering\epsfxsize=3.6in \epsfysize=3.6in \epsffile{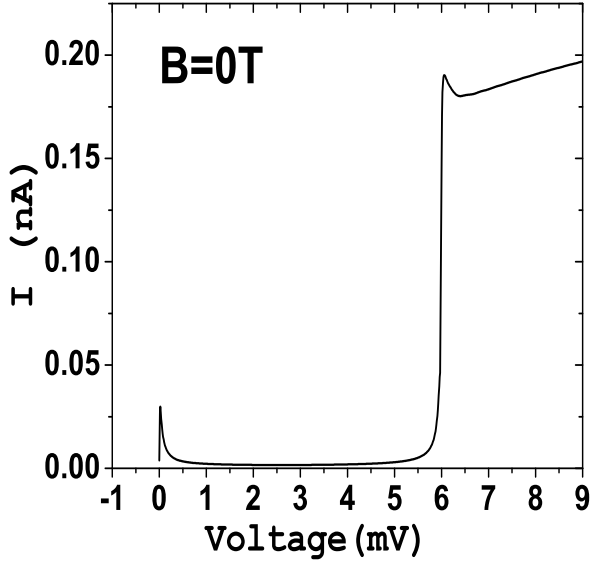}
\caption{Stationary $I/V_{DC}$ (B=0). At low $V_{DC}$, $I$ takes
place when one electron from the (1,1) spin-singlet, tunnels to the
double occupied singlet state in the right QD (0,2). At higher
$V_{DC}$ the system reaches the states
$|3\rangle=|\uparrow,\uparrow\rangle$ and
$|4\rangle=|\downarrow,\downarrow\rangle$ (inter-dot triplet states)
and $I$ drops off due to SB. The SB region is the plateau between
the two main peaks, where a finite current leakage is observed due
to spin-flip induced by Hyperfine interaction. At larger $V_{DC}$
($V_{DC}\geq 6$ meV) the chemical potential of the right lead
crosses the inter-dot triplet state and the right QD becomes
suddenly discharged producing a large peak in $I$.}
\end{figure}
\begin{figure}
\centering\epsfxsize=3.4in \epsfysize=3.5in\epsffile{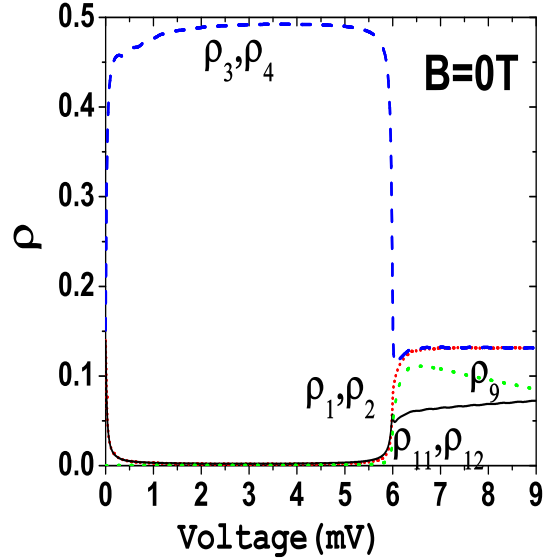}
\caption{(Color on line). States occupations versus $V_{DC}$ for
$B=0$. In the SB region the occupation probabilities are dominated
by states $|3\rangle=|\uparrow,\uparrow\rangle$ and
$|4\rangle=|\downarrow,\downarrow\rangle$. See blue (dashed) line.
At larger $V_{DC}$ ($V_{DC}\geq 6$ meV) the chemical potential of
the drain crosses the inter-dot triplet state and the right QD
becomes suddenly discharged. This produces a large peak in the
current and a dramatic reduction in the occupation of states
$|3\rangle$ and $|4\rangle$ as expected. At the same time other
states, which are now important in the transport, increase their
occupations.}
\end{figure}

Because of the mismatch between nuclear and electronic Zeeman
energies  spin-flip transitions must be accompanied at low
temperature by phonon emission\cite{erlingsson2001}. Phonon
absorption is not possible at low temperature. Therefore for $B\neq
0$ Hyperfine interaction only produces electronic spin-flip
$relaxation$ processes. We approximate the current-limiting
spin-flip transition rate from parallel-spin to opposite-spin
configurations by:
\begin{equation}
\left[\frac{1}{\tau_{sf}}\right]_{L(R)} \simeq\frac{2\pi}{\hbar} |<
\hat{H}_{sf}>|^{2}
%\frac{\gamma}{\Delta Z_{e}^{2}+\gamma^{2}} \label{eq:spinflip}
\frac{\gamma}{(\Delta Z_{e})_{L(R)}^{2}+\gamma^{2}}
\label{eq:spinflip}
\end{equation}
where the width $\gamma$ is the electronic state life-time
broadening which is of the order of $\mu eV$ ($\gamma\simeq 5 \mu
eV$), i.e., of the order of the phonon scattering rate
\cite{fujisawa}. This equation shows that a different number of
nuclei or different splitting Zeeman can give rise to a different
spin-flip rate in each dot.

The splitting Zeeman is given by:
\begin{equation}
   (\Delta Z_{e})_{L(R)}=g_{e}\mu_{B} B +\frac{A}{2}P_{L(R)}
\end{equation}
 is the total electronic Zeeman splitting including the Overhauser shift produced
by the effective nuclear B.
\begin{equation}
   (\Delta Z_{Overhauser})_{L(R)}=\frac{A}{2}P_{L(R)}
\end{equation}
We assume that a weakly coupled QD's model do no consider molecular
states. The basis of states considered reflects this situation. As a
consequence, the exchange coupling constant is zero and the Zeeman
splitting is given by equation (14).

The expressions we propose for the the electronic spin-flip
scattering rate $W_{i,j}^{sf}$ depend on the different processes:
\begin{eqnarray}
|\downarrow,\downarrow\rangle\rightarrow|\uparrow,\downarrow\rangle&\Rightarrow&
W_{5,4}^{sf} = \left[\frac{1}{\tau_{sf}}\right]_{L} \left[\frac{1+P_L}{2}\right]\\
|\downarrow,\downarrow\rangle\rightarrow|\downarrow,\uparrow\rangle&\Rightarrow&
W_{6,4}^{sf} = \left[\frac{1}{\tau_{sf}}\right]_{R} \left[\frac{1+P_R}{2}\right]\\
|\uparrow,\uparrow\rangle\rightarrow|\uparrow,\downarrow\rangle&\Rightarrow&
W_{5,3}^{sf} = \left[\frac{1}{\tau_{sf}}\right]_{R} \left[\frac{1-P_R}{2}\right]\\
|\uparrow,\uparrow\rangle\rightarrow|\downarrow,\uparrow\rangle&\Rightarrow&
W_{6,3}^{sf}= \left[\frac{1}{\tau_{sf}}\right]_{L}
\left[\frac{1-P_L}{2}\right]
\end{eqnarray}

The equations that describe the time evolution of the nuclei spin
polarization for both dots include the flip-flop interaction and a
phenomenological nuclear spin relaxation time $\tau_{relax}\approx
100 s$ \cite{ono2} for the scattering between nuclei:
\begin{eqnarray}
\dot P_{L} &=&W_{6,3}^{sf}\rho_{3}-W_{5,4}^{sf}\rho_4
-\frac{P_L}{\tau_{relax}}\\
\dot P_{R} &=&W_{5,3}\rho_{3}-W_{6,4}^{sf}\rho_4-
\frac{P_R}{\tau_{relax}}
\end{eqnarray}

Including spin-flip interactions, the rate equation for the
occupation probability of $|3\rangle=|\uparrow,\uparrow\rangle$ and
$|4\rangle=|\downarrow,\downarrow\rangle$ is:
\begin{eqnarray}
&&\dot{\rho}_{3}=W_{3,1}\rho_{1}+W_{3,7}\rho_{7}+W_{3,11}\rho_{11}\nonumber\\
&&-\left(W_{1,3}+W_{7,3}+W_{11,3}+W^{sf}_{5,3}+W^{sf}_{6,3}\right)\rho_{3}\\
&&\dot{\rho}_{4}=W_{4,2}\rho_{2}+W_{4,8}\rho_{8}+W_{04,12}\rho_{12}\nonumber\\
&&-\left(W_{2,4}+W_{8,4}+W_{12,04}+W^{sf}_{5,4}+W^{sf}_{6,4}\right)\rho_{3}
\end{eqnarray}

\subsection{Total current expression}
The system of time evolution equations for the electronic states
occupations $ \rho_i $ and nuclei polarization of the left and right
dot is self-consistently solved. From that we calculate the total
current through the system which is the physical observable of
interest. The current going from the left lead to the left QD is
defined as:
\begin{eqnarray}
&&I_L=I^{\uparrow}_L + I^{\downarrow}_L=\nonumber\\
&&e\{(W_{3,1}\rho_1+W_{5,2}\rho_2+W_{14,12}\rho_{12} +
W_{16,13}\rho_{13})\nonumber\\
&&-(W_{1,3}\rho_3+W_{2,5}\rho_5+
W_{12,14}\rho_{14}+W_{13,16}\rho_{16})\}^{\uparrow} \nonumber\\
&& + e\{(W_{6,1}\rho_1+
W_{4,2}\rho_2+W_{17,12}\rho_{12}+W_{15,13}\rho_{13})\nonumber\\
 &&-(W_{1,6}\rho_6+W_{2,4}\rho_4+W_{12,17}\rho_{17}+
W_{13,15}\rho_{15})\}^{\downarrow}\nonumber\\
\end{eqnarray}
where the first (second) bracket, $\{\}^{\uparrow}
(\{\}^{\downarrow})$, represents the up (down) current. Similar
expressions can be obtained for the inter-dot current ($I_{12}$) and
for the current going from the right dot to the collector ($I_{R}$).
In general the total current through the device is:
\begin{equation}
 I=\frac{I_{L}+I_{1,2}+I_{R}}{3}
\end{equation}

\section{III. Results}
Experimental results\cite{ono} show, for magnetic field $B$=0, a
peak at low $V_{DC}$, a big plateau and a peak of high intensity at
large $V_{DC}$, ($V_{DC}\geq 6$ meV). We present similar calculated
results in Fig. 2. At low $V_{DC}$, $I$ takes place when one
electron from the (1,1) spin-singlet, tunnels to the double occupied
singlet state in the right QD (0,2). This process happens when the
extra energy to add one electron to the right QD is given by a
nearby gate voltage. At higher $V_{DC}$ the system reaches the
states $|3\rangle=|\uparrow,\uparrow\rangle$ and
$|4\rangle=|\downarrow,\downarrow\rangle$ (inter-dot triplet states)
and $I$ drops off due to SB. It corresponds to the observed plateau.
This can be observed in Fig. 3, where the state occupation is
presented versus applied bias. According to it, the SB region is
governed by the states $|3\rangle$ and $|4\rangle$ with an
occupation of almost $0.5$ for each state. In this $V_{DC}$ region
the only way for one electron in the left QD to tunnel to the right
one would be through an excited state in the right QD. Nevertheless
at zero $B$, is too high in energy to participate in the transport
window. Despite SB, there is a finite leakage current measured in
the SB region due to the finite probability for electrons in the
QD's to flip their spin by interaction with nuclei. At larger
$V_{DC}$ ($V_{DC}\geq 6$ meV) the chemical potential of the right
lead crosses the inter-dot triplet state and the right QD becomes
suddenly discharged producing a large peak in the current.  This
peak is mainly due, according to our calculations, to the inelastic
contribution to the current with the emission of acoustic phonons.
The elastic contribution is rather small because the ground levels
in each dot involved in the tunneling
:$|\uparrow,0\rangle\rightarrow|0,\uparrow\rangle$ and
$|\downarrow,0\rangle\rightarrow|0,\downarrow\rangle$, are totally
out of resonance. This situation corresponds very well with the
schematic diagram in the Fig. 1a.

\begin{figure}
\centering\epsfxsize=3.4in \epsfysize=4.0in\epsffile{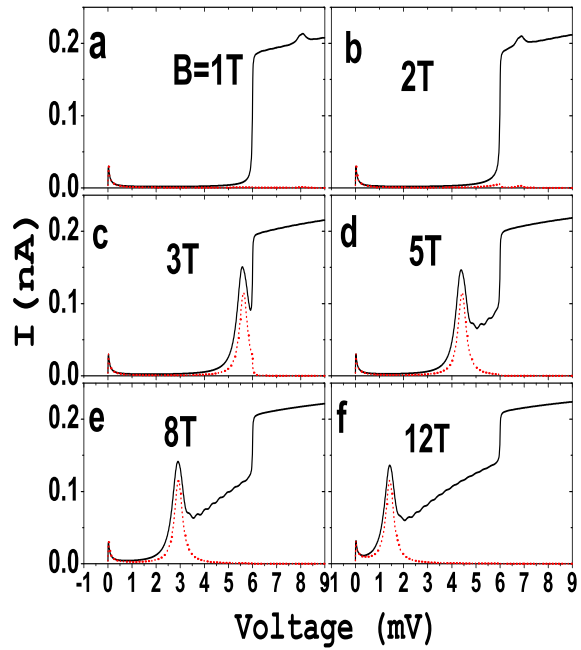}
\caption{(Color on line). Stationary $I/V_{DC}$ curve calculated for
different $B$. We observe an additional peak at finite $B$ which
moves to lower $V_{DC}$ as $B$ increases. For different values of
$B$ the resonance condition: $|3\rangle=|\uparrow,\uparrow\rangle
\Rightarrow |7\rangle = |0,\uparrow{\uparrow}^*\rangle$ and
$|4\rangle=|\downarrow,\downarrow\rangle \rightarrow |8\rangle =
|0,\downarrow{\downarrow}^*\rangle$ occurs at different values of
$V_{DC}$. A shoulder at the right side of the $B$-dependent peak is
also observed. The single line corresponds to both elastic
transitions (direct tunneling) and inelastic transitions
(phonon-assisted contributions). The dotted line means that only
elastic transitions are taken into account. In this case it can be
observed that the shoulder and the large right peak collapses. This
demonstrates that inelastic transitions play a crucial role to
sustain the current not only in the large right peak, but also in
the right shoulder of the moving central peak. Again elastic
inter-dot transitions do not contribute because the corresponding
levels in each dot are out of resonance. The resonant peaks move to
lower $V_{DC}$ as $B$ increases as expected. For larger $B$ the
energy shift for the excited state is also larger and its final
position is getting closer to the ground state. As a result a lesser
$V_{DC}$ is required to get the resonant condition. }
\end{figure}

A finite $B$ parallel to the current, produces an energy shift
experienced by the Fock-Darwin states due to its coupling with the
electronic orbital momentum. $B$ couples with the electronic orbital
angular momentum of the first excited state of the right QD ($l=1$).
An additional shift in the electronic energy comes as well from the
additional confinement potential induced by $B$ \cite{leo}. These
$B$-dependent shifts and the typical energy scales of this problem
(on-site Coulomb energy or orbital level spacing) are much larger
than the Zeeman splitting. Thus, Zeeman splitting is neglected only
in terms of tunneling processes. Increasing $B$ the first excited
state of the right QD (0, $\uparrow\uparrow^*$) enters in the
transport window and comes into resonance with the ground state of
the left QD. This opens a new transport channel, and thus $I$ flows
through the device and SB is lifted. In Fig. 4 we present the
stationary $I/V_{DC}$ curve calculated for different $B$. We observe
an additional peak at finite $B$ which moves to lower $V_{DC}$ as
$B$ increases. For different values of $B$, the resonance condition
%$|3\rangle=|\uparrow,\uparrow\rangle \rightarrow|7\rangle =
%|0,\uparrow{\uparrow}^*\rangle$ and
%$|4\rangle=|\downarrow,\downarrow\rangle \rightarrow |8\rangle =
%|0,\downarrow{\downarrow}^*\rangle$,
occurs at different values of
$V_{DC}$. In Fig. 4, the single line corresponds to both elastic
(direct tunneling) and inelastic (phonon-assisted) contributions.
The dotted line represents only elastic inter-dot transitions.
According to these results, inelastic transitions play a crucial
role to sustain the current. They are responsible of the large right
peak and the right shoulder of the moving central peak. In the
latter, elastic inter-dot transitions do not contribute because the
corresponding levels in each dot are out of resonance (See fig. 1b).
The results presented in Fig. 4 are in good agreement with the
experimental curve by Ono et al.\cite{ono}, including the shoulder
at the right side of the $B$-dependent peak. The oscillations
however are smeared out in the experiment. The reason being that we
did not include in our model damping of the bosonic system
corresponding to phonon cavity losses\cite{brandesprb}. It has been
shown that the system is extremely sensitive even to very small
damping\cite{brandesprb}.

\begin{figure}
\centering\epsfxsize=3.4in \epsfysize=5.0in\epsffile{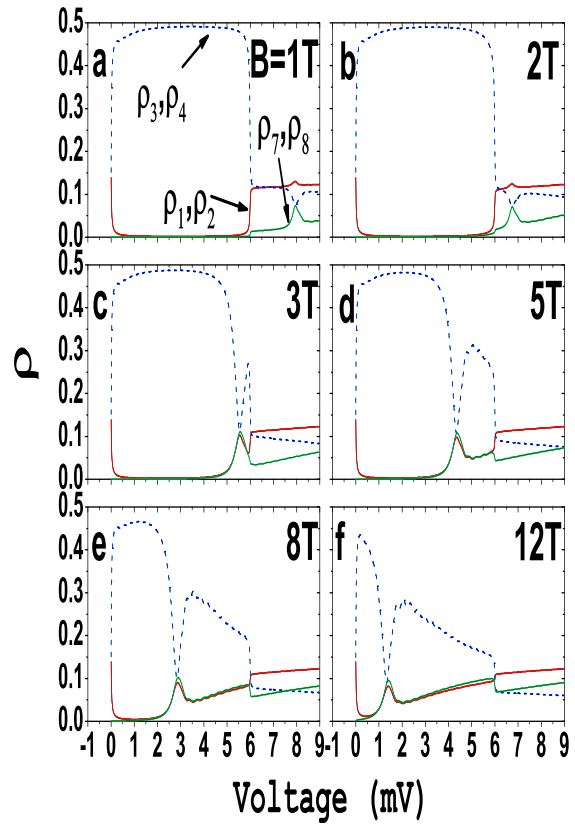}
\caption{(Color on line). Stationary charge occupation of the
electronic states for the same cases of $B$ as in Fig. 4. In the SB
region the inter-dot triplet states($|3\rangle$ and $|4\rangle$) are
occupied in the QD's, for lower values of $B$. For larger $B$ the
resonant condition between the ground state of the left QD and the
excited state of the right QD is fulfilled inside the SB region.
This give rise to spin blockade removal. The opening of this new
current channel
 decreases $\rho_3$ and $\rho_4$ and increases $\rho_7$ and $\rho_8$}
\end{figure}
\begin{figure}
\centering\epsfxsize=3.6in \epsfysize=4.5in\epsffile{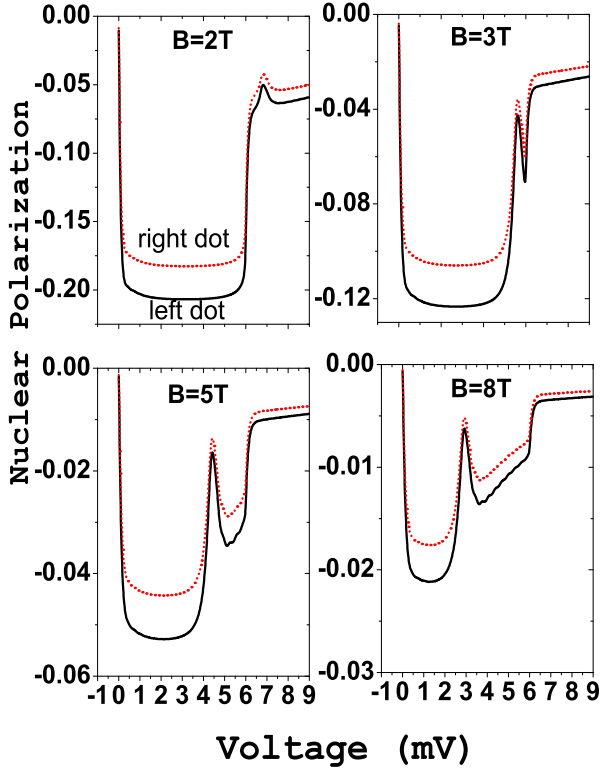}
\caption{(Color on line). Nuclear polarizations of left and right
dot versus applied voltage for four different $B$ separately
$B=2,3,5$ and $8T$. For $B\neq 0$ only electronic spin-flip
relaxation process are possible: from spin down to spin up. For the
nuclei we have the opposite process: from spin up to down. This
explains the negative polarization obtained for both dots. The
variation of $P_L(R)$ is opposite to the electronic occupation of
state $|4\rangle$ as expected. The decreasing polarization magnitude
versus increasing $B$ is a consequence of the splitting Zeeman (see
eqn. 13)). }
\end{figure}
\begin{figure}
\centering\epsfxsize=3.6in \epsfysize=4.5in\epsffile{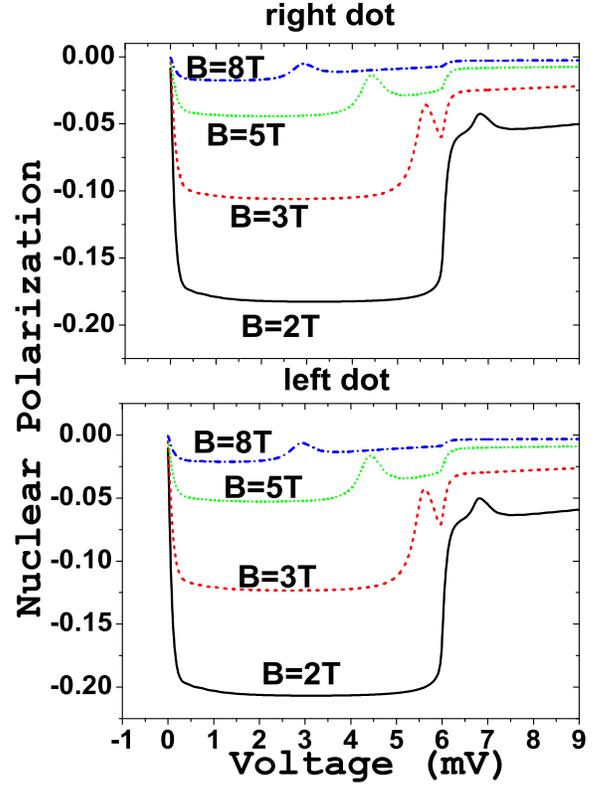}
\caption{(Color on line). Nuclear polarization for both dots and all
magnetic fields jointly. The negative values of $P_L(R)$ decrease as
the magnetic field increases. This happens in both dots. If we
increase $B$, $\Delta Z_{e}$ increases and consequently spin-flip
rate and nuclear polarization decrease as expected.}
\end{figure}
Fig. 5 shows the charge occupation for different states and for the
same values of $B$ as in Fig. 4. For small $B$ the electrons occupy
mainly the inter-dot triplet states ($|\uparrow,\uparrow\rangle$ and
$|\downarrow,\downarrow\rangle$) in the SB region with a probability
of almost 0.5 for each one. No SB removal is observed apart from a
finite leakage current due to sf by HF interaction. In this case the
resonant condition between the ground state of the left dot and the
first excited state of the right dot ($|\uparrow,\uparrow\rangle
\Rightarrow |0,\uparrow{\uparrow}^*\rangle$ or
$|\downarrow,\downarrow\rangle \Rightarrow
|0,\downarrow{\downarrow}^*\rangle$), is fulfilled at $V_{DC}>6mV$.
However at larger $B$ the resonant condition happens inside the SB
region giving rise to blockade lifting. The opening of this new
current channel corresponds to a decrease in the occupation of
$|3\rangle$ and $|4\rangle$ and to an increase in the occupation of
$|7\rangle$ and $|8\rangle$.  The shoulder at the right side of the
resonant peak produces also a removal of SB. This is due to
inelastic tunneling through the inner barrier assisted with the
emission of acoustic phonons. Summarizing, SB removal is produced by
three processes: the first one is electronic spin-flip by HF
interaction. The second one is  elastic tunneling through the right
dot excited state coming into the transport window by $B$. Finally,
additional inelastic contributions to the inner tunneling assisted
by emission of acoustic phonons.

The microscopical model for the Hyperfine interaction, allows us to
study  the dynamical nuclear polarization in each dot\cite{levitov}.
We calculate the nuclear polarizations $P_{L(R)}$, versus the
applied voltage and different magnetic fields. We have considered
that $N_{L}=10^{6}$ and $N_{R}=1.1\times10^{6}$. As a consequence we
obtain different spin-flip rates for each dot. We present the
obtained results in Figs. 6 and 7. In Fig. 6 we represent the
nuclear polarizations of left and right dot versus applied voltage
for four different $B$ separately $B=2,3,5$ and $8T$. As we said
above, for $B\neq 0$ only electronic spin-flip relaxation process
are possible: from spin down to spin up. This means that for nuclei
we have the opposite process, i.e., from spin up to spin down that
explains the negative nuclear polarization. For all magnetic fields
studied, the right dot presents an smaller negative polarization.
This is because of the larger number of nuclei in the right dot (see
eqn. (12)). The peculiar shape that these graphs present can be
explained if we rewrite the dynamical equations of the nuclear
polarization of left and right dot. Now we have to take into account
that only electronic spin-flip relaxation processes are allowed and
that the nuclear spin relaxation time is very large. Finally the two
equations read:
\begin{eqnarray}
\dot P_{L} &=&-W_{5,4}^{sf}\rho_4\\
\dot P_{R} &=&-W_{6,4}^{sf}\rho_4
\end{eqnarray}

According to these expressions the calculated values for $P_L(R)$
versus applied voltage are opposite to ones obtained for $\rho_4$.
That means that when the occupations of state $|4\rangle$ increases
(SB region) nuclear polarization should increase (in negative) too.
However, outside of the SB region or when the SB is removed (new
transport channel through state
$\rho_8=|0,\downarrow{\downarrow}^*\rangle$) nuclear polarization
decreases because $\rho_4$ is much smaller.

In Fig. 7, we present in two panels the nuclear polarization for
both dots and all magnetic fields jointly. The negative values of
$P_L(R)$ decrease as the magnetic field increases. This happens in
both dots. The explanation comes readily if we observe the
expression of the spin-flip scattering rate (eqn. 13). The highest
values for this rate corresponds to an splitting Zeeman $\Delta
Z_{e}\simeq0$, that is obtained at $B=0$. In this situation spin
down and spin up electronic states are degenerate. However if we
increase $B$, $\Delta Z_{e}$ increases and consequently spin-flip
rate and nuclear polarization decrease. Summarizing, we obtain a
dynamical nuclear polarization that is intimately related with the
electronic occupation through the device.

\section{IV. Conclusions}
In conclusion we reproduce experimental features by Ono et
al.\cite{ono}, and show how, inelastic transitions with the
corresponding emission of phonons play a crucial role sustaining the
current through the device. We show as well that the interplay
between the electron and nuclei spin distributions within the dots
are responsible for lifting the SB yielding a leakage current.  We
demonstrate also that at finite $B$, the participation of excited
states in the current lifts SB. We develop a microscopical model to
account for the Hyperfine interaction in each dot. Using this model
we study the dynamical nuclear polarization as a function of the
applied voltage. We obtain that nuclear polarization is closely
related with the electronic occupations. Our results indicate that a
combination of $B$ and $V_{DC}$ allows to control the current
through the device, making these systems potential components for
spintronics and quantum computing.

\section{Acknowledgments}
This work has been supported by the MCYT (Spain) under grant
MAT2005-06444 (JI and GP), by the Ram\'on y Cajal program (J.I.) by
the EU Human Potential Programme: HPRN-CT-2000-00144, by the Welch
Foundation (AHM) and by the DOE (AHM) under grant DE-FG03-02ER45958.
%%%%%%%%%%%%%%%%%%%%%%%%%%%%%%%%%%%%%%%%%%%%%%%%%%%%%%%%%%%%%%%%%%
%%%%%%%%%
%BIBLIOGRAPHY

\end{document}